# Co-evolution between Codon Usage and Protein-Protein Interaction in Bacteria

Maddalena Dilucca [1,*], Giulio Cimini [2,3], Sergio Forcelloni[1] and Andrea Giansanti [1,4]

[1] Dipartimento di Fisica, Sapienza University of Rome, Italy
[2] Dipartimento di Fisica and INFN Roma2 unit, Tor Vergata University of Rome, Italy
[3] Istituto dei Sistemi Complessi (CNR) UoS Sapienza University of Rome, Italy
[4] INFN Roma1 unit, Sapienza University of Rome, Italy
[*] Correspondence: maddalena.dilucca@gmail.com

**Abstract:** We study the correlation between the codon usage bias of genetic sequences and the network features of protein-protein interaction (PPI) in bacterial species. We use PCA techniques in the space of codon bias indices to show that genes with similar patterns of codon usage have a significantly higher probability that their encoded proteins are functionally connected and interacting. Importantly, this signal emerges when multiple aspects of codon bias are taken into account at the same time. The present study extends our previous observations on *E.Coli* over a wide set of 34 bacteria. These findings could allow for future investigations on the possible effects of codon bias on the topology of the PPI network, with the aim of improving existing bioinformatics methods for predicting protein interactions.

## 1. Introduction

The systematic analysis of protein-protein interaction (PPI) is key to understand the patterns of chemical reactions within the cell, as well as the role played by proteins in regulative processes [1]. On the applicative side, comparing the interactomes of different species may allow understanding disease-related processes that engage more than one species, such as host-pathogen relationships, to identify clinically relevant host-pathogen PPI, and consequently developing future therapeutic applications [2,3].

An important aspect to take into account when studying PPI is the degeneracy of the genetic code, due to the presence of synonymous codons at the genetic level that encode the same amino acid in the translated protein. Although synonymous codons are indistinguishable in the primary structure of a protein, they are not used randomly, but with different frequencies that may vary across species, across regions of the same genome, and even across regions of the same gene. This phenomenon, known as Codon Usage Bias (CUB) [4-6], is well-established in the literature, despite a general understanding of its biology still lacks [7]. It is known however that CUB is involved in many important cellular processes, including differential gene expression [8-10], translation efficiency and accuracy [11], gene function and dynamics of the ribosome [12,13], co-translational folding of the proteins [14], and deamination of tRNA anticodons [15]. CUB is believed to be maintained by a balance between mutation-selection (random variability in genetic sequences followed by fixation of the optimal codons) and genetic drift (allowing for the occurrence of non-optimal codons) [16]. Indeed, highly expressed genes feature a strong CUB by using a small subset of codons, optimized by translational selection, while the presence of non-optimal codons in less-expressed genes causes long breaks during protein synthesis that affect the folding process [17]. Furthermore, CUB is well structured along the genome, with neighbor genes having similar usage frequencies of synonymous codons [18].

Considering that gene co-expression level and proximity between the positions of the genes in the genome are powerful predictors of protein-protein interaction [19,20], it would be interesting to analyze how the similarity in CUB of the genes is reflected into the likelihood that the corresponding proteins make physical contact in the cell. Given the above considerations, we would expect a more similar codon usage bias between interacting proteins than non-interacting ones. Recent evidence in this direction has been provided in some case studies for *E. Coli* and yeast. [21,22]. In [23] we showed that in *E.Coli* translational selection systematically favors optimal codons in proteins that have a large number of interactors and belong to the most representative communities in the PPI. In the present work our aim is precisely to understand whether the similarity of codon usage patterns between a pair of genes is related in general to the possible interaction of the corresponding proteins.

By extending the analysis in [23] to a large set of unrelated bacterial species, here we provide basic observations of sufficient generality on the co-evolution of CUB and the connectivity features of bacterial interactomes. Specifically, our main result indicates that the functional structuring of the PPI network has interfered with the peculiar codon choice of the genes over evolution. Our findings point out that CUB should be a relevant parameter in the prediction of unknown protein-protein interactions from genomic information.

## 2. Materials and Methods

### 2.1 Genomic Sequences

In this work, we select a set of 34 bacterial genomes with different behavior, environment and taxonomy (see Table 1 for details). Each bacterium represents a specific clade in the phylogenetic tree by Plata et al. [24]. Nucleotide sequences were downloaded from the FTP server of the National Center for Biotechnology Information (ftp://ftp.ncbi.nlm.nih.gov/genomes/archive/old_genbank/Bacteria/) [25].

**Table 1.** Summary of the 34 bacterial datasets considered in this work. For each specie we report the organism name, abbreviation, RefSeq, STRING code, size of genome (number of genes *n*), and density of the PPI network — defined as ratio between the number of links in the real interactome and the maximum number of possible links, namely $n(n-1)/2$, where *n* is the number of proteins.

| Organisms | Abbr. | RefSeq | STRING | Size | Density |
|---|---|---|---|---|---|
| Agrobacterium fabrum str. C58 | agtu | NC_003062 | 176299 | 2765 | 0.008 |
| Anabaena variabilis ATCC 29413 | anva | NC_007413 | 240292 | 5043 | 0.005 |
| Aquifex aeolicus VF5 | aqae | NC_000918 | 224324 | 1497 | 0.009 |
| Bifidobacterium longum NCC2705 | bilo | NC_004307 | 216816 | 1726 | 0.004 |
| Bordetella bronchiseptica RB50 | bobr | NC_002927 | 257310 | 4994 | 0.005 |
| Bordetella parapertussis 12822 | bopa | NC_002928 | 360910 | 4185 | 0.008 |
| Brucella melitensis bv. 1 str. 16M | brme | NC_003317 | 224914 | 2059 | 0.006 |
| Buchnera aphidicola str. Bp | buap | NC_004545 | 224915 | 504 | 0.008 |
| Burkholderia pseudomallei K96243 | bups | NC_006350 | 272560 | 3398 | 0.002 |
| Buchnera aphidicola Sg uid57913 | busg | NC_004061 | 198804 | 546 | 0.002 |
| Burkholderia thailandensis E264 | buth | NC_007651 | 271848 | 3276 | 0.001 |
| Caulobacter crescentus | cacr | NC_011916 | 565050 | 3885 | 0.002 |
| Campylobacter jejuni | caje | NC_002163 | 192222 | 1572 | 0.004 |
| Corynebacterium efficiens YS-314 | coef | NC_004369 | 196164 | 2938 | 0.006 |
| Corynebacterium glutamicum ATCC 13032 | cogl | NC_003450 | | 2959 | 0.005 |
| Chlamydia trachomatis D/UW-3/CX | chtr | NC_000117.1 | 272561 | 894 | 0.008 |
| Clostridium acetobutylicum ATCC 824 | clac | NC_003030.1 | 272562 | 3602 | 0.005 |
| Francisella novicida U112 | frno | NC_008601 | 401614 | 1719 | 0.007 |
| Fusobacterium nucleatum ATCC 25586 | funu | NC_003454.1 | 190304 | 1983 | 0.002 |
| Haemophilus ducreyi 35000HP | hadu | NC_002940 | 233412 | 1717 | 0.004 |
| Klebsiella pneumoniae | klpn | NC_009648 | 272620 | 4775 | 0.005 |
| Listeria monocytogenes EGD | limo | NC_003210 | 169963 | 2867 | 0.003 |
| Mesorhizobium loti MAFF303099 | melo | NC_002678.2 | 266835 | 6743 | 0.0001 |
| Mycoplasma genitalium G37 | myge | NC_000908 | 243273 | 475 | 0.005 |
| Mycoplasma pneumoniae M129 | mypn | NC_000912.1 | 272634 | 648 | 0.006 |
| Mycobacterium tuberculosis H37Rv | mytu | NC_000962.3 | 83332 | 3936 | 0.006 |
| Porphyromonas gingivalis ATCC 33277 | pogi | NC_010729 | 431947 | 2089 | 0.001 |
| Ralstonia solanacearum GMI1000 | raso | NC_003295.1 | 267608 | 3436 | 0.002 |
| Sphingomonas wittichii RW1 | spwi | NC_009511 | 392499 | 4850 | 0.007 |
| Staphylococcus aureus NCTC 8325 | stau | NC_007795 | 93061 | 2767 | 0.004 |
| Synechocystis sp. PCC 6803 | sysp | NC_000911.1 | 1148 | 3179 | 0.004 |
| Thermotoga maritima MSB8 | thma | NC_000853.1 | 243274 | 1858 | 0.001 |
| Vibrio cholerae N16961 | vich | NC_002505 | 243277 | 2534 | 0.001 |
| Xylella fastidiosa 9a5c | xyfa | NC_002488 | 160492 | 2766 | 0.002 |

**2.2 Codon Usage Bias Measures**

In the last years, different metrics to measure CUB have been proposed. In this work we use the following four indices to characterize a genetic sequence (we remand to [6,23] for the detailed definitions). 1) The Relative Synonymous Codon usage (RSCU) of a codon is the number of occurrences of that codon in the genome, with respect to the family of synonymous codon it belongs to. RSCU values can be combined into the Effective Number of Codons (NC) [26], which is a popular statistical measure of the number of codons used in a sequence. 2) The tRNA Adaptation index (tAI) [27] is instead a widely used metric based on gene expression levels, which builds on the assumptions that tRNA availability is the driving force for translational selection. 3) CompAI and CompAI_w [23] are two recently proposed metrics that refine tAI by using the competition between cognate and near-cognate tRNA to proxy the efficiency of codon-anticodon coupling. 4) The GC content of a gene, namely the percentage of guanine and cytosine in the RNA molecules, is a parameter used to explain CUB differences between species [28].

**2.3 Protein-Protein Interaction Network**

The PPI networks of the 34 bacterial genomes were retrieved from the STRING database (Known and Predicted Protein-Protein Interactions) [29]. Given that a predicted interaction in STRING is assigned with a confidence level $w$, as typically done in PPI studies we select as actual links of the networks only those interactions with $w>0.9$. The resulting degree (namely the number of incident link) of a protein is denoted as $k$.

To detect the communities of a PPI we use the Molecular Complex Detection (MCODE) method [30]. MCODE works by iteratively grouping together neighboring nodes with similar values of the core-clustering coefficient, which is defined as the density of the highest $k$-core of its immediate neighborhood times $k$ (here a $k$-core is a sub-network of minimal degree $k$). Thus, MCODE detects the densest regions of the network and assigns a score to each community equal to its size times its internal link density. In line with our previous study [23], here we consider only the first eight MOCDE communities.

## 2.4 Principal Component Analysis

Principal Component Analysis (PCA) [31] is a multivariate statistical method that transforms a set of possibly correlated variables into a set of linearly uncorrelated ones (called principal components, spanning a space of lower dimensionality). The transformation is defined so that the first principal component accounts for the largest possible variance of the data, and each succeeding component in turn has the highest variance possible under the constraint that it is orthogonal to (i.e., uncorrelated with) the preceding components.

We use PCA over the space of the five codon bias indices described above. Thus, for a given species, each gene is represented as a 5-dimensional vector with coordinates (compAI, compAI_w, tAI, NC, GC). These coordinates are separately normalized to zero mean and unit variance over the genome of the species. The principal components are then the eigenvectors of the covariance matrix, ordered according to the magnitude of the corresponding eigenvalues.

## 2.5 Null Network Model and Statistical Tests

For a given species, in order to characterize the CUB patterns over the interactome we have to compare the PPI network with a suitable null network model, which should embody a null hypothesis of no relation between the codon usage of two genes and the possible interactions between their encoded proteins.

Here we use the *configuration model* (CM), namely a degree-preserving randomization of the network links which thus destroys the original structure of the network (see [32] for an introduction to the method). Note that by constraining the degrees, the model automatically takes care of the linking bias for highly connected proteins, which typically corresponds to essential genes [33] (but also to genes that are conserved across species or related to ribosomal functions [34]).

Once the CM is built, we can assess the significance of a given set of link-related quantities by comparing their distribution on the original PPI network with their distribution on the null model. The Mann-Whitney U test is used to determine if the two distributions are different (we use a p-value threshold of $10^{-3}$). Alternatively, to assess the significance of a single network quantity $X$, we use the Z-score $Z[X] = (X - \langle X \rangle)/\sigma_X$ where $\langle X \rangle$ and $\sigma_X$ are its mean and standard deviation computed in the null model. Thus, the Z-score quantifies the number of standard deviations by which the actual and null model values of $X$ differ.

## 3. Results and Discussion

### 3.1 Interacting proteins do not share a common codon usage statistic nor tRNA adaptation level

As mentioned in the introduction, our aim is to analyze how the closeness in codon usage of two genes is reflected in the capacity of their proteins to make physical contact in the cell, and we expect a more similar CUB between interacting proteins than non-interacting ones.

We analyze one species at a time. For a given species, we start by characterizing each gene by its 61-component vector of RSCU values, which provides the detailed statistics of codon usage in the sequence. We can then quantify how similar are two genes in the use of synonymous codons through the normalized scalar product of their RSCU vectors. We thus compute the distribution of the scalar products between the RSCU vectors of the genes whose encoded proteins are linked in the PPI. We compare this distribution with the analogous distribution computed on the null model network. Table 2 reports the p-values of the Mann-Whitney U test between these distribution for all species in the dataset. We notice that many species do not pass the test (having a p-values larger than the threshold $10^{-3}$), in which cases we can conclude that the two distributions are statistically equal. Therefore, CUB is not very predictive of protein interactions when measured only through codon usage statistics, without taking into account the information about tRNA levels.

We can perform the same exercise using the (normalized) difference in tAI levels (rather than the scalar product of RSCU) in order to qualitatively assess whether similarity of tRNA abundance and adaptation, without sequences statistic, can explain protein connectivity. Results of the Mann-Whitney U test reported in Table 2 show that tAI (used as a proxy of gene expressivity) is rarely informative about PPI connections, and in general less informative than codon usage statistics.

Before moving to the next section, two remarks are in order. Firstly, we do not test the difference in CompAI index because its distribution on the various interactomes turn out to be too narrow for the Mann-Whitney U test to work properly, and we also do not test GC since it is not a direct measure of CUB but rather a contributing factor reflecting mutational bias [35] Secondly, rather than gene expressivity it would be much more interesting to test gene co-expressivity, which is known to have significant correlation to PPIs. However, gene co-expression data are available only for a handful of species, and thus can be employed in specific case studies but not for a species-wide assessment.

**Table 2.** P-values of to the Mann-Whitney U test, for the pairwise comparisons between the normalized distribution of RSCU scalar products and tAI differences for genes corresponding to interacting proteins in the PPI, and their distribution obtained in the randomized CM of the PPI. For each species, we report the organism name, abbreviation, and p-value of RSCU and tAI statistics. In bold we report statistically significant values.

| Organisms | Abbr. | p-val RSCU | p-val tAI |
|---|---|---|---|
| Agrobacterium tumefaciens | agtu | **$1.3*10^{-6}$** | **$1.1*10^{-3}$** |
| Anabaena variabilis ATCC 29413 | anva | **$1.0*10^{-4}$** | **$1.2*10^{-4}$** |
| Aquifex aeolicus VF5 | aqae | $2.0*10^{-2}$ | **$1.2*10^{-5}$** |
| Bifidobacterium longum NCC2705 | bilo | **$1.5*10^{-3}$** | $8.9*10^{-1}$ |
| Bordetella bronchiseptica RB50 | bobr | **$2.7*10^{-3}$** | $8.7*10^{-1}$ |
| Bordetella parapertussis 12822 | bopa | **$1.1*10^{-3}$** | $1.2*10^{-2}$ |
| Brucella melitensis bv. 1 str. 16M | brme | **$1.6*10^{-7}$** | $1.3*10^{-2}$ |
| Buchnera aphidicola str. Bp | buap | **$1.3*10^{-3}$** | $8.7*10^{-1}$ |
| Burkholderia pseudomallei K96243 | bups | **$1.2*10^{-6}$** | **$1.2*10^{-5}$** |
| Buchnera aphidicola Sg uid57913 | busg | $3.5*10^{-2}$ | $5.0*10^{-2}$ |
| Burkholderia thailandensis E264 | buth | **$9.1*10^{-15}$** | $4.0*10^{-2}$ |
| Caulobacter crescentus | cacr | $1.5*10^{-1}$ | $3.0*10^{-2}$ |
| Campylobacter jejuni | caje | $1.0*10^{-1}$ | $1.2*10^{-1}$ |
| Corynebacterium efficiens YS-314 | coef | **$2.7*10^{-3}$** | $8.9*10^{-1}$ |
| Corynebacterium glutamicum ATCC 13032 | cogl | **$6.0*10^{-3}$** | $1.4*10^{-1}$ |
| Chlamydia trachomatis D/UW-3/CX | chtr | $1.8*10^{-1}$ | $2.0*10^{-2}$ |
| Clostridium acetobutylicum ATCC 824 | clac | **$7.0*10^{-11}$** | $5.0*10^{-2}$ |
| Francisella novicida U112 | frno | $6.6*10^{-1}$ | $7.0*10^{-2}$ |
| Fusobacterium nucleatum ATCC 25586 | funu | **$1.1*10^{-11}$** | **$1.9*10^{-6}$** |
| Haemophilus ducreyi 35000HP | hadu | **$5.7*10^{-3}$** | **$1.1*10^{-5}$** |
| Klebsiella pneumoniae | klpn | **$8.9*10^{-3}$** | **$3.0*10^{-4}$** |
| Listeria monocytogenes EGD | limo | **$1.3*10^{-5}$** | **$2.0*10^{-3}$** |
| Mesorhizobium loti MAFF303099 | melo | $1.0*10^{-2}$ | $2.4*10^{-1}$ |

| Mycoplasma genitalium G37 | myge | **$5.8*10^{-5}$** | $5.4*10^{-1}$ |
| Mycoplasma pneumoniae M129 | mypn | **$5.2*10^{-5}$** | $7.8*10^{-2}$ |
| Mycobacterium tuberculosis H37Rv | mytu | $1.6*10^{-2}$ | $4.5*10^{-2}$ |
| Porphyromonas gingivalis ATCC 33277 | pogi | **$4.5*10^{-6}$** | $5.0*10^{-3}$ |
| Ralstonia solanacearum GMI1000 | raso | $2.2*10^{-2}$ | **$5.0*10^{-3}$** |
| Sphingomonas wittichii RW1 | spwi | **$8.9*10^{-9}$** | **$4.0*10^{-4}$** |
| Staphylococcus aureus NCTC 8325 | stau | $2.0*10^{-2}$ | $2.3*10^{-1}$ |
| Synechocystis sp. PCC 6803 | sysp | **$3.0*10^{-4}$** | $2.0*10^{-2}$ |
| Thermotoga maritima MSB8 | thma | **$1.6*10^{-3}$** | **$1.5*10^{-4}$** |
| Vibrio cholerae N16961 | vich | **$2.1*10^{-8}$** | $3.4*10^{-1}$ |
| Xylella fastidiosa 9a5c | xyfa | **$5.1*10^{-4}$** | $2.2*10^{-1}$ |

## 3.2 Principal Component Analysis over the space of CUB indices

A possible way to obtain a more evident correlation between CUB similarity and PPI connectivity is to combine the information coming from the various facets of codon bias, namely codon usage statistics, mutational selection, tRNA expression levels and coupling efficiency, respectively measured by NC, GC, tAI and CompAI. Thus, for a given species, we then perform PCA over the space of the five codon bias indices (CompAI, CompAI_w, tAI, NC and GC content) measured separately for each gene in the genome (see Figure 1 for an example; plots for all species are shown in the Supplementary Materials). Typically, the first and second principal components ($PC_1$ and $PC_2$) turn out to represent for as much as 65% of the total variance of codon bias over the genome. Additionally, projection of these two principal components on the individual CUB indices (loadings) shows that none of the five indices predominantly contributes to the data variability. We can thus focus on the plane defined by the $PC_1$ and $PC_2$ vectors (see Figure 2 for an example; plots for all species are shown in the Supplementary Materials), where the placement of a gene depends on a weighted contribution of all the CUB indices.

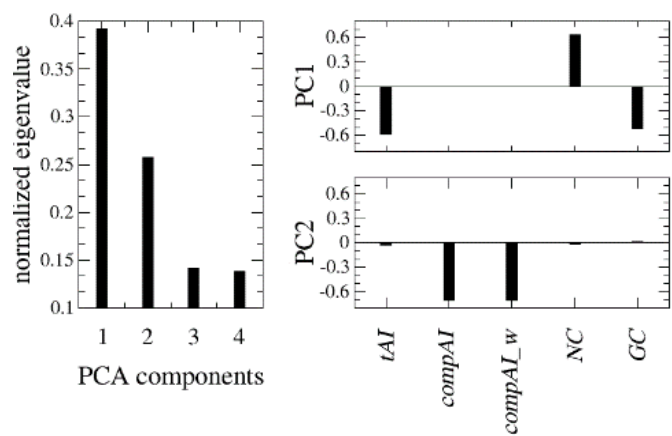

**Figure 1.** PCA results for the example *agtu* species. (Left panel) Eigenvalues of the PCA analysis, showing the first and second principal components ($PC_1 - PC_2$) turn out to represent as much as 65% of the total CUB variance. (Right panels) Projection of these two components over the space of CUB indices. The other bacteria species are shown in the Supplementary Materials.

We can also place on this plane the centroids of the top eight MCODE communities of the PPI network, where coordinates and error bars of each centroid are obtained as the coordinate mean and standard deviations of the genes belonging to the respective module. The first community (composed overall by 97 % of genes belonging to COG class J, related to translation, ribosomal structure and biogenesis) is typically well separated from the others. Concerning the other communities, the situation depends on the species (see Figure 2 of Supplementary Materials): some bacteria such as *caje*, *chtr* and *pogi* do not have separated centroids, whereas many other bacteria such as *bups, buth* and *myge* have all eight communities well separated and localized. In these latter cases we can conclude that when a set of proteins are physically and functionally connected in a module, then their corresponding genes tend to share common codon bias features. This observation could be explained by considering that interacting proteins (especially those belonging to the same community) need to be present in the cell according to precise quantities at a given time to form the protein complexes required for the ongoing cellular programs.

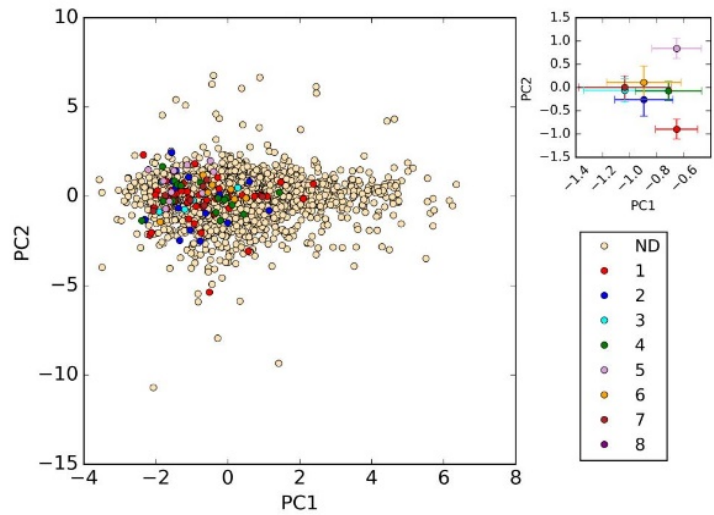

**Figure 2.** Representation of each gene in the $PC_1 - PC_2$ plane, for the example *agtu* species. The inset shows the centroids of the top-eight MCODE communities, with error bars denoting the standard deviation of the distribution of points around the centroid. The other bacteria are shown in the Supplementary Materials.

**3.3 Z-score profiles: The closer the codon usage of genes, the higher the probability of protein interaction**

A last, we use the Euclidean distance $d$ between two genes on the $PC_1 - PC_2$ plane as a proxy of their overall codon usage similarity. We can then compute, for each species, the conditional probability $Pr(link|d)$ of a physical or functional pair interaction between proteins, given that their coding genes fall within a distance $d$ in the plane of the two principal PCA components. In other words, $Pr(link|d)$ is the fraction of gene pairs, among those localized within a distance $d$, whose encoded proteins are connected in the PPI network. In order to obtain a statistically significant profile, we compare $Pr(link|d)$ estimated on the real interactome with $Pr(\langle link \rangle|d)$, namely the same probability estimated on the configuration model (CM) of the network. We recall that CM is used as the null hypothesis that no relation exists between the codon usage of two genes and the interaction between the encoded proteins. The significance of $Pr(link|d)$ with respect to the null hypothesis is thus quantified through the Z-score

$$Z_d = \frac{\Pr(link|d) - \Pr(\langle link \rangle|d)}{\sigma[\langle link \rangle|d]}$$

Figure 3 shows the Z-score as a function of the gene distance *d* for some example bacterial species (plots for all species are shown in the Supplementary Materials). Interestingly, a typical pattern emerges. For small distances ($d \leq 3$), the probability of finding a connection between two proteins in the empirical interactome is significantly much higher than in the null model. Conversely, for larger distances ($d > 3$) the real PPI and the CM become statistically compatible (and sometimes, for $3 < d < 5$, links are even less likely than in the null model). This pattern is evident for all the 24 bacteria that pass the test for the RSCU distributions (see Table 2), although not significantly in four cases (*anva*, *bopa*, *coef*, *vich*). Notably, the same pattern is observed also for eight bacteria (*busg*, *cacr*, *caje*, *chtr*, *frno*, *klpn*, *raso*, *stau*, *mytu*) that instead do not pass the RSCU test. In contrast, only two bacteria (*aqae* and *melo*) are characterized by a different Z-score profile (for *melo* this is probably due to its low PPI density). We can thus conclude that, as a general rule, the distance between a pair of genes in the plane of the first two PCA components is a statistically robust predictor of the likelihood that their corresponding proteins interact (physically or functionally). In agreement with our previous results [23], the signal is more evident when codon usage frequencies of interacting proteins are far from being random.

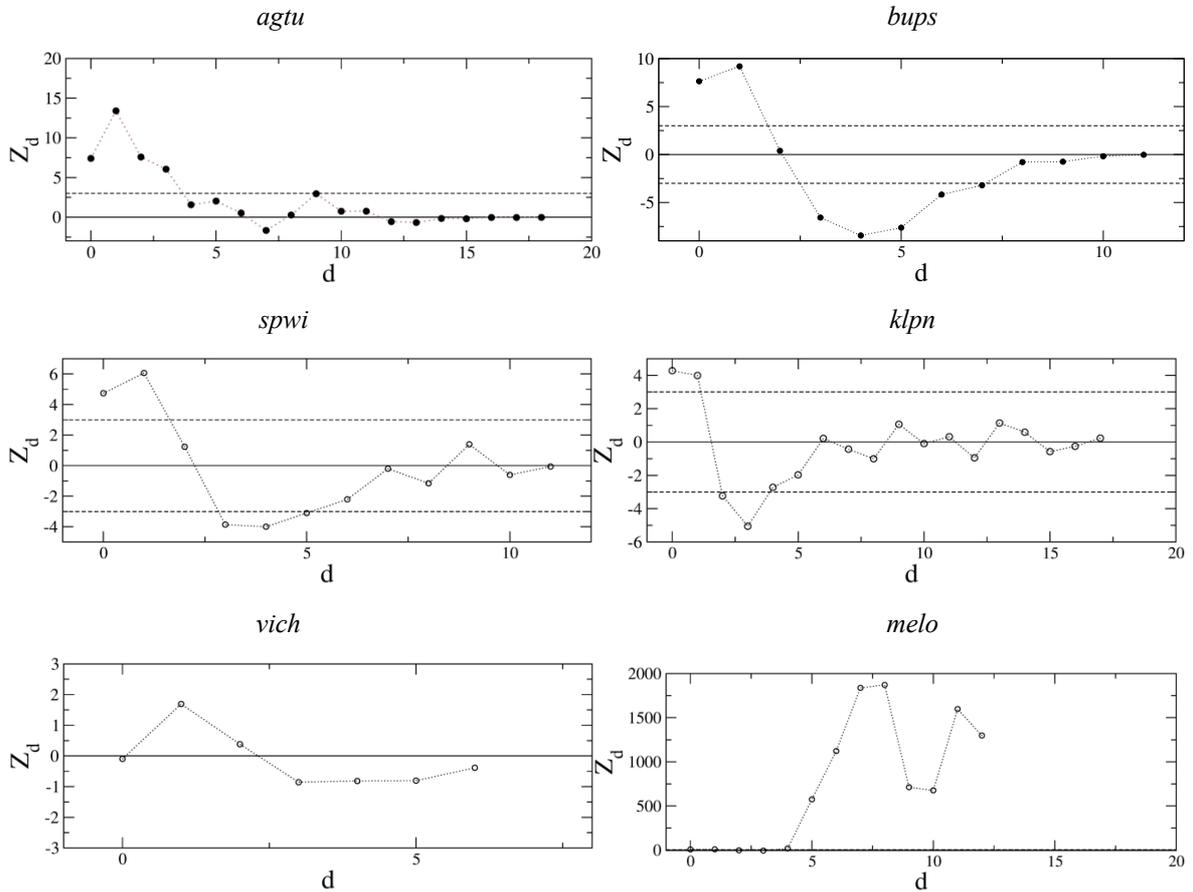

**Figure 3.** Z-scores of $Pr(link|d)$ as a function of the Euclidean distance *d* between the codon usage bias of pair of genes (computed via PCA). The horizontal dashed lines mark the significance interval of ±3 standard deviations. We show a few example species. The other bacteria are shown in the Supplementary Materials.

## 4. Conclusion

In this work we studied how the coherence in codon usage among genes is reflected in the capacity of their encoded proteins to interact in the protein network. For this purpose, we have extended our previous work on the case study of *E. Coli* [23] to a set of other 34 bacterial genomes characterized by different taxonomy [24]. As a general rule, we find that CUB as measured solely by either the occurrence frequencies of synonymous codons or tRNA abundance levels is not much able to distinguish between proteins that make contacts or not in the PPI network. Conversely, by combining the different facets of CUB (as expressed by NC, tAI, CompAI, GC), we observe that highly connected proteins belonging to the same communities in the protein interaction network are encoded by genes that are coherent in their codon choices. Specifically, our results provide evidence that if two genes have similar codon usage patterns, then the corresponding proteins have a significant probability of being functionally connected or physically interacting. Consequently, this study provides new information based on the similarity in codon usage of genes that can be potentially integrated into existing computational prediction methods of protein-protein interaction. Additionally, as recent studies point out (see for instance [36]), using CUB as an additional level of information in the study of protein interaction networks could be useful to identify genes linked to infections, drug-resistance or altered metabolism, and thus hint at alternative treatments in the light of growing resistance to antibiotics and the propagation of infectious agents [37].

**References**

1. Gavin CKW, Porras P, Aranda B, Hermjakob H, and Orchard SE, Analyzing protein–protein interaction networks, J. Proteome Res. 2012, 11, 4, 2014-2031
2. Shah PS, Wojcechowskyj JA, Eckhardt M, Krogan NJ, Comparative mapping of host–pathogen protein–protein interactions. Curr Opin Microbiol. 2015; 27: 62–68
3. Arnold R, Boonen K, Sun MGF, Kim PM, Computational analysis of interactomes: Current and future perspectives for bioinformatics approaches to model the host–pathogen interaction space. Methods 2012; 57(4): 508-518
4. Wright F, The effective number of codons' used in a gene. Gene. 1990;87(1):23-29
5. Behura SK, Severson DW, Codon usage bias: causative factors, quantification methods and genome-wide patterns: with emphasis on insect genomes, Biol Rev Camb Philos Soc. 2013; 88(1): 49-61
6. Hanson G and Coller J, Codon optimality, bias and usage in translation and mRNA decay, Nat Rev Mol Cell Biol. 2018, 19(1): 20-3
7. Tuller T, Challenges and obstacles related to solving the codon bias riddles, Biochem Soc Trans. 2014; 42(1):155-159
8. Gouy M, Gautier C. Codon usage in bacteria: correlation with gene expressivity. Nucleic Acids Res. 1982; 10(22):7055–7074
9. Quax TE, Claassens NJ, Soll D, van der Oost J, Codon bias as a means to fine-tune gene expression, Mol Cell. 2015; 59(2):149-61
10. Hunter B. Fraser, Aaron E. Hirsh, Dennis P. Wall, Michael B. Eisen, Coevolution of gene expression among interacting proteins, PNAS 2004, 101 (24) 9033-9038
11. Sabi R, Tuller T., Modelling the efficiency of codon-tRNA interactions based on codon usage bias. DNA Res. 2014; 21(5):511-26
12. Najafabadi HS, Goodarzi H, Salavati R, Universal function-specificity of codon usage, Nucleic Acids Res. 2009; 37(21):7014-23
13. Dilucca M, Cimini G, Giansanti A, Essentiality, conservation, evolutionary pressure and codon bias in bacterial genes, Gene 2018; 663, 178-188
14. Zhao F, Yu Ch, Liu Y, Codon usage regulates protein structure and function by affecting translation elongation speed in Drosophila cells, Nucleic Acids Res 2017; 45(14), 8484-8492
15. Rafels-Ybern A, Torres AG, Grau-Bove X, Ruiz-Trillo I, de Pouplana LR, Codon adaptation to tRNAs with Inosine modification at position 34 is widespread among Eukaryotes and present in two Bacterial phyla, RNA Biol 2018; 15(4-5):500-507
16. Kober KM, Pogson GH, Genome-wide patterns of codon bias are shaped by natural selection in the purple sea urchin, strongylocentrotus purpuratus, G3: Genes, Genomes, Genetics 2013, 3 (7), 1069-1083

17. Pop C, Rouskin S, Ingolia NT, Han L, Phizicky EM, Weissman JS. Causal signals between codon bias, mRNA structure, and the efficiency of translation and elongation. Mol Syst Biol. 2014; 10:770
18. Plotkin JB, Kudla G, Synonimous but not the same: the causes and consequences of codon bias, Nat. Rev. Genet 2011; 12(1):32-42
19. Jansen R, Greenbaum D, Gerstein M, Relating whole-genome expression data with protein-protein interactions, Genome Res 2002; 12(1):37-46
20. Fraser HB, Hirsh AE, Wall DP, Eisen MB, Coevolution of gene expression among interacting proteins, Proc Natl Acad Sci USA 2004; 101(24):9033-8
21. Najafabadi HS, Salavati R, Sequence-based prediction of protein-protein interactions by means of codon usage. Genome Biol. 2008; 9(5):R87
22. Zhou Y, Zhou YS, He F, Song J, Zhang Z. Can simple codon pair usage predict protein–protein interaction? Molecular BioSystems 2012; 8(5):1396-1404
23. Dilucca M, Cimini G, Semmoloni A, Deiana A, Giansanti A, Codon bias patterns of E. coli's interacting proteins. PLoS ONE 2015; 10(11): e0142127
24. Plata G, Henry CS, Vitkup D, Long-term phenotypic evolution of bacteria, Nature 2015; 517, 369-372
25. Benson DA, Karsch-Mizrachi I, Clark K, Lipman DJ, Ostell J, Sayers EW, GenBank, Nucleic Acids Res. 2012;40 (Database issue):D48-D53
26. Fuglsang A, The 'effective number of codons' revisited, Biochem. Biophys. Res. Commun. 2004; 317 (3): 957-964
27. dos Reis M, Savva R, Wernisch L., Solving the riddle of codon usage preferences: a test for translational selection, Nucleic Acids Res. 2014; 32(17):5036-5044
28. Hershberg R, Petrov DA, Selection on codon bias. Annual Review of Genetics. 2008; 42: 287-99
29. Szklarczyk D, Franceschini A, Wyder S, von Mering C et al. STRING v10: Protein-protein interaction networks, integrated over the tree of life, Nucleic Acids Res 2015; 43(Database issue):D447-52
30. Bader GD, Hogue CWV. An automated method for finding molecular complexes in large protein interaction networks. BMC Bioinformatics. 2003;4:2
31. Jolliffe IT, Principal component analysis, Second Edition, Springer (2002)
32. Cimini G, Squartini T, Saracco F, Garlaschelli D, Gabrielli A, Caldarelli G. The statistical physics of real-world networks. Nat Rev Phys 2019; 1(1), 58-71
33. Jeong H, Mason SP, Barabasi AL, Oltvai ZN. Lethality and centrality in protein networks. Nature. 2001;411(6833):41–2
34. Dilucca M, Cimini G, Giansanti A. Bacterial protein interaction networks: connectivity is ruled by gene conservation, essentiality and function. arXiv:1708.02299
35. Li J, Zhou J, Wu Y, Yang S, Tian D, GC-Content of synonymous codons profoundly influences amino acid usage. G3 (Bethesda). 2015; 5(10): 2027–2036
36. Rajkumari J, Chakraborty S, Pandey P, Distinctive features gleaned from the comparative genomes analysis of clinical and non-clinical isolates of Klebsiella pneumoniae, Bioinformation 2020; 16(3):256-268
37. Zoragh R, Reiner NE, Protein interaction networks as starting points to identify novel antimicrobial drug targets. Current Opinion in Microbiology 2013; 16(5): 566-572

**Figure S1.** Eigenvalues of the PCA analysis, and projection of $PC_1$ and $PC_2$ over the five CUB indices. Each plot corresponds to a specie in our dataset.

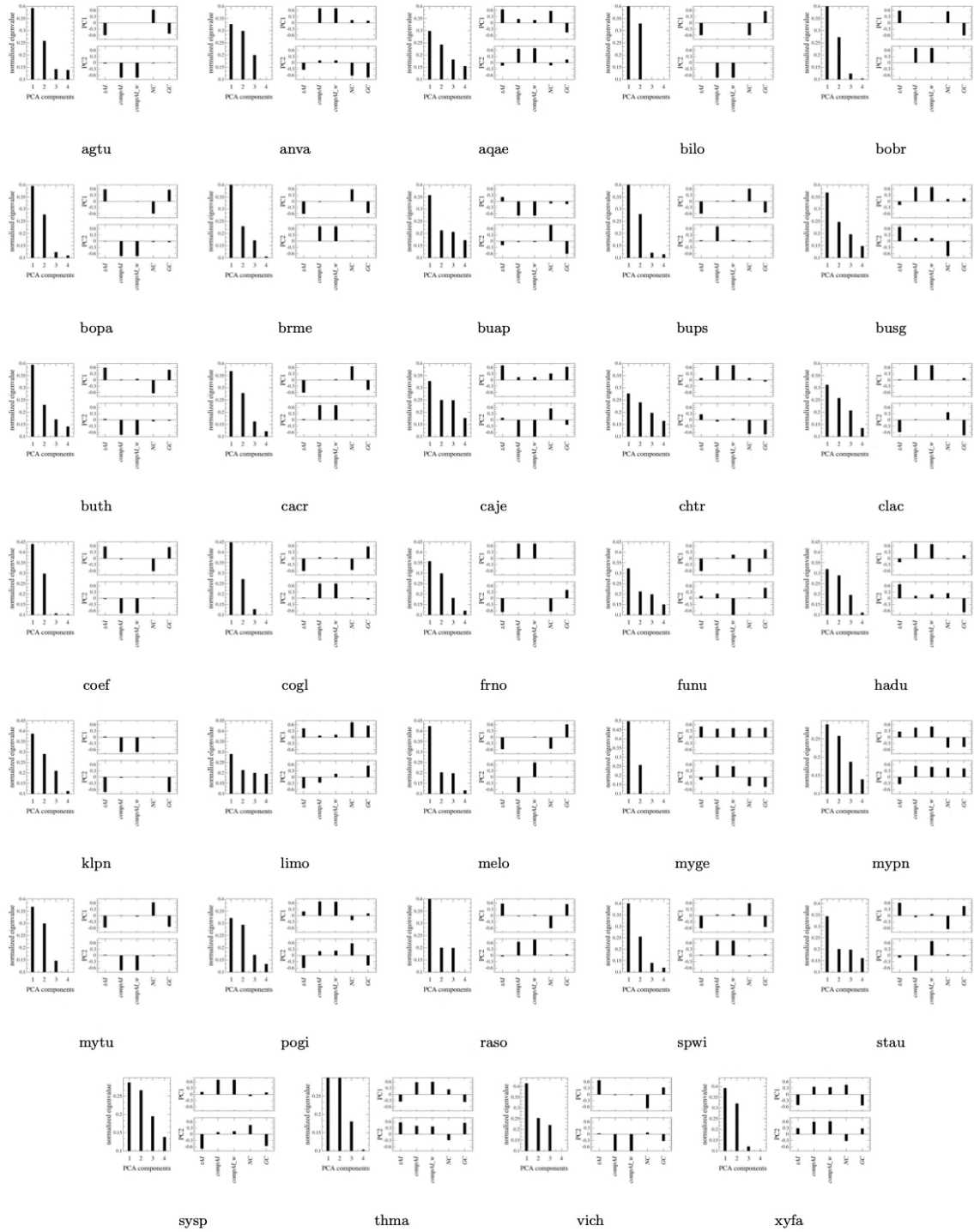

**Figure S2.** Representation of each gene in the $PC_1 - PC_2$ plane. The inset reports the centroids of the top-eight MCODE communities, with error bars denoting the standard deviation of the distribution of points around the centroid. Each plot corresponds to a specie in our dataset.

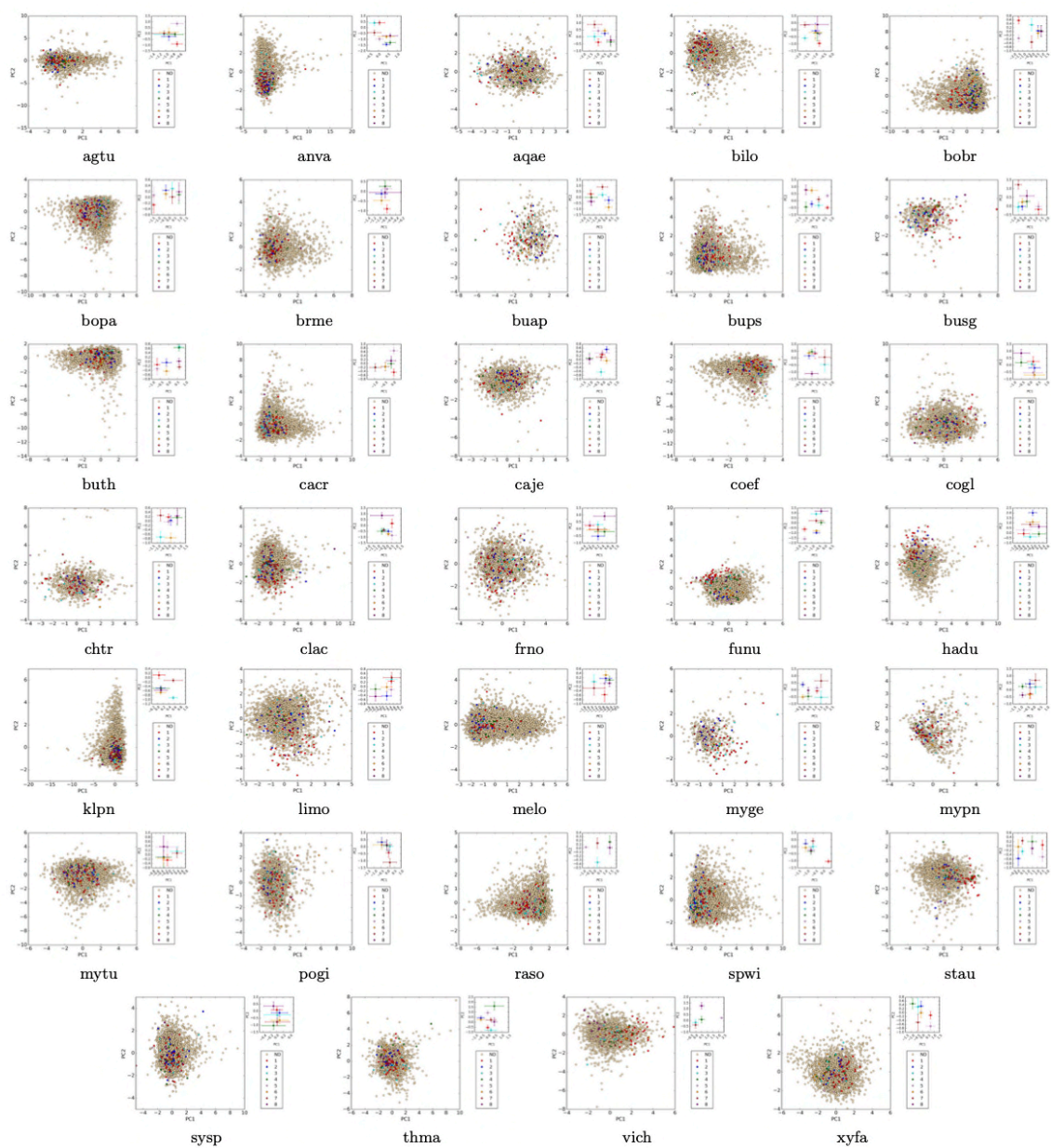

**Figure S3.** Z-scores of $Pr(link|d)$ as a function of the Euclidean distance $d$ between the codon usage bias of pair of genes (computed via PCA). The horizontal dashed lines mark the significance interval of ±3 standard deviations. Each plot corresponds to a specie in our dataset.

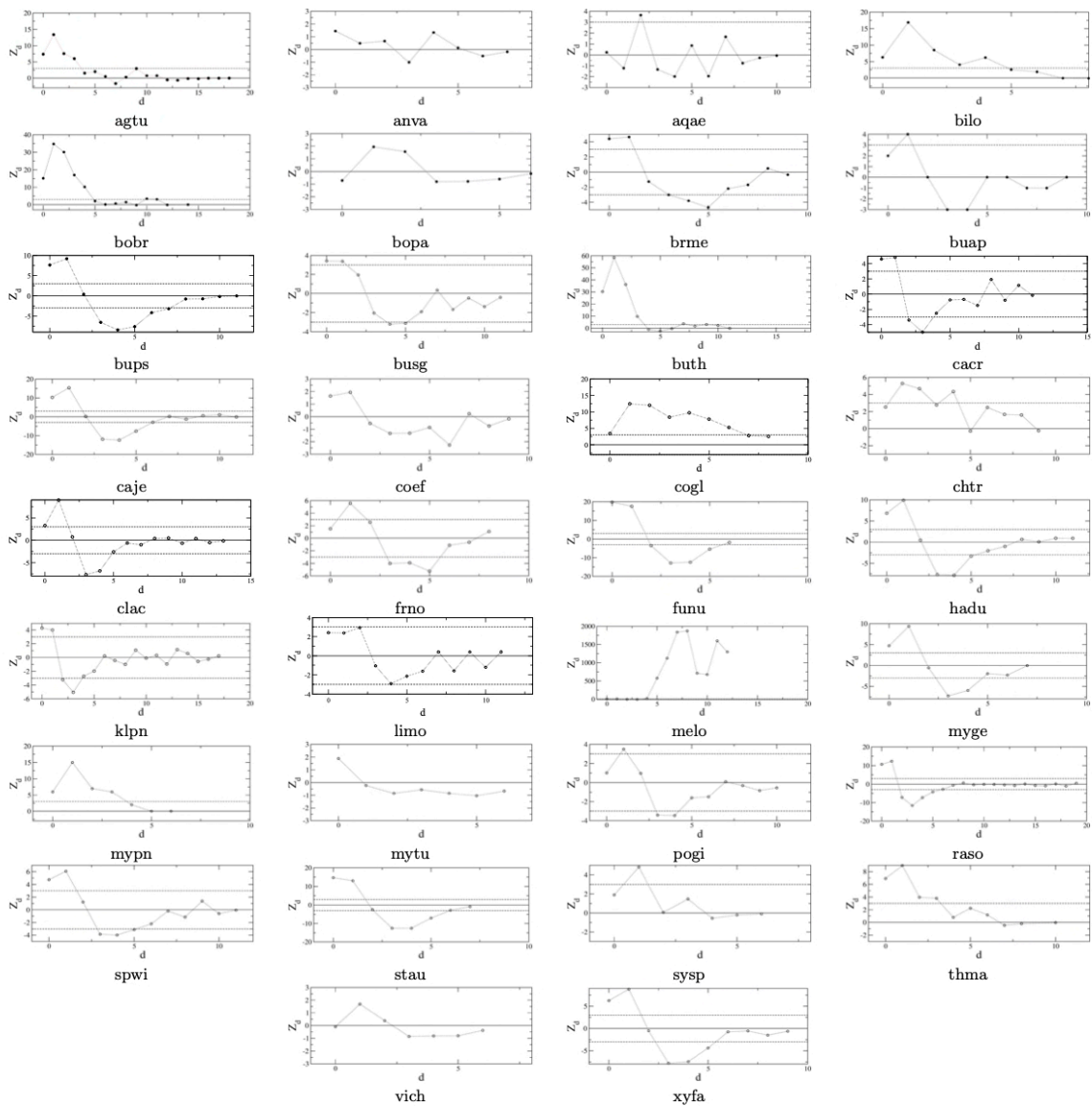